\documentclass{jpp}
\usepackage{graphicx}
\usepackage[utf8]{inputenc}
\usepackage[T1]{fontenc}
\usepackage{amsmath}
\usepackage{xcolor}
\usepackage{changes}

\shorttitle{Ambipolar effect self-consistently produced}
\shortauthor{L. Barbieri}
\title{Self-consistent generation of the ambipolar electric field in collisionless plasmas via multi-mode electrostatics}
\author{Luca Barbieri\aff{1}
  \corresp{\email{luca.barbieri@obspm.fr}}
 }
\affiliation{\aff{1} LIRA, Observatoire de Paris, Université PSL, Sorbonne Université, Université Paris Cité, CY Cergy Paris Université, CNRS, 92190
Meudon, France
}

\begin{document}
\maketitle

\begin{abstract}
%To address the long-standing coronal heating problem, a novel kinetic model of a two-component plasma atmosphere has been recently introduced, successfully reproducing the observed temperature and density profiles of the solar atmosphere and low-mass main-sequence stars \citep{Barbieri2023temperature,Barbieri2024b,Barbieri2025}. In this framework, the coronal plasma particles are subjected to solar gravity, the Pannekoek-Rosseland electric field, and self-electrostatic interactions treated within the simplified Hamiltonian mean-field (HMF) model. A key limitation of this approach is the complete neglect of collisional effects in the coronal plasma. However, it is well known that collisions significantly alter the electric field structure, rendering the Pannekoek-Rosseland field insufficient to maintain charge neutrality. Instead, a self-generated ambipolar electric field arises. In this paper, we demonstrate how the ambipolar effect can be self-consistently recovered within the model by extending the electrostatic interaction to include multiple Fourier modes.
In this work, we investigate the generation of the ambipolar electric field in a gravitationally stratified, collisionless plasma atmosphere. In such environments, gravity tends to separate charged species. To prevent separation an electric field, classically described by the Pannekoek–Rosseland expression, is usually imposed externally. Here, we propose a self-consistent method to recover this field based on a multi-mode Fourier expansion of the electrostatic interaction. We show that, under suitable conditions, this approach naturally leads to the ambipolar electric field and restores charge neutrality. The method is tested in both isothermal and multi-temperature plasma configurations. This framework provides a foundation for future developments that may include collisions, ionization, and asymmetric boundary conditions to model more realistic stellar atmospheres.
\end{abstract}
%%%%%%%%%%%%%%%%%%%%%%%%%%%%%%%%%%%%%%%%%%%%%%%%%%%%
\keywords{plasma dynamics, plasma properties, plasma nonlinear phenomena}
%%%%%%%%%%%%%%%%%%%%%%%%%%%%%%%%%%%%%%%%%%%%%%%%%%%%%%%%%%%%%%%%%%%%%%

\maketitle
\section{Introduction}\label{sec1}

In the solar corona, the outermost layer of the solar atmosphere, the temperature rises from thousands to millions of degrees with increasing altitude, while the density simultaneously drops by more than two orders of magnitude. Understanding how the coronal plasma achieves this stationary configuration remains a fundamental and unresolved issue, commonly known as the coronal heating problem \citep{Parker:1972wu,Ionson_1978,Heyvaerts_Priest_1983,Scudder1992a,Scudder1992b,Dmitruk:1997uf,2005ApJ...618.1020G,Rappazzo:2008vl,Pontieu:2011vg,2013ApJ...773L...2R,2015RSPTA.37340265W,2020A&A...636A..40H,Hau_2025}.

The upper chromosphere and the base of the transition region are known to be highly dynamic, characterized by intense, short-lived, and small-scale brightenings
\citep{Dere:1989ux, Teriaca:2004wy, Peter:2014uz,Tiwari:2019us, Berghmans:2021wl}.

Prompted by these observational evidences, recently a new kinetic model of the solar atmosphere has been proposed \citep{Barbieri2023temperature,Barbieri2024b}, demonstrating, through both numerical and analytical techniques, that suitable, rapid, and random temperature perturbations in the upper chromosphere can drive the overlying plasma toward a stationary configuration exhibiting the observed inverted temperature and density profiles (often referred to as temperature inversion).

In this model, the coronal plasma is represented as a one-dimensional, collisionless, two-species system subject to constant downward gravity and an electric field (the Pannekoek-Rosseland field) generated by the Sun to ensure charge neutrality \citep{Pannekoek_1922,Rosseland_1924}. As pointed out in \cite{Barbieri2023temperature,Barbieri2024b,Barbieri2025}, the most significant weakness of this approach lies in its omission of Coulomb collisions, despite the low Knudsen number ($K_n\sim 10^{-2}-10^{-3}$ in the transition region and $K_n\sim 10^{-1}$ in the corona), which indicates that collisions are non-negligible in both regions.

As shown by \cite{Landi-Pantellini2001} via numerical simulations, Coulomb collisions modify the electric field in such a way that the standard Pannekoek-Rosseland field is no longer sufficient to maintain charge neutrality. Instead, a supplementary contribution from the plasma's self-consistent electrostatic field is required. In \cite{Barbieri2023temperature,Barbieri2024b}, electrostatic interactions were approximated using the Hamiltonian Mean-Field (HMF) model \citep{AntoniRuffo:pre1995,Chavanis2005,Elskens2019}, which truncates the Fourier expansion of the electrostatic potential at the first mode.

Therefore, in order to introduce collisions into the model, it is first necessary to understand whether the HMF-based modeling developed in this work is capable of reproducing the ambipolar effect. For this reason, we do not include collisions in the present study where we focus on the collisionless limit. Our aim is to determine what form of the self-generated electric field is required to induce the ambipolar effect that leads to the Pannekoek–Rosseland field, without imposing it externally as done in previous works \citep{Barbieri2023temperature,Barbieri2024b,Barbieri2025}. We will show that by retaining an increasing number of Fourier modes in the electrostatic potential, it is possible to recover the ambipolar effect and thus the Pannekoek–Rosseland field in a fully self-consistent manner. This framework provides the basis for including ambipolar effects in plasma atmospheres and will support future extensions of the model, such as the inclusion of collisions.

The structure of the paper is as follows. Section 2 revisits the inadequacy of the Pannekoek-Rosseland field in ensuring charge neutrality in a collisional plasma out of thermal equilibrium. Section 3 introduces the coronal loop model from \citep{Barbieri2023temperature,Barbieri2024b}, extended to include multiple electric field modes. Section 4 shows both theoretically and numerically how the ambipolar electric field can be generated by this approach. Finally, Section 5 summarizes our findings and outlines potential future directions.

\section{Charge neutrality in a gravitationally stratified stellar atmosphere}
We consider a plane-parallel atmosphere composed of a plasma of electrons and protons stratified by the gravitational field. For each species, the hydrostatic equilibrium condition is expressed as
\begin{equation}\label{pressureequations}
    \frac{d P_{\alpha}}{dz} = n_{\alpha}F_{\alpha} +R_{\alpha}, \quad F_{\alpha}=\mathrm{sign}(e_{\alpha}) e E -g m_{\alpha} \quad,
\end{equation}
$\alpha \in \{e,p\}$ denotes the species (electrons or protons), $m_{\alpha}$ is the mass of a particle of species $\alpha$, $e$  is the elementary charge, $g$ is the gravitational acceleration, $z$ is the height, $n_{\alpha}$ is the number density, $P_{\alpha}$ the pressure, $F_{\alpha}$ the total force per unit volume, $E$  the electric field, and $R_{\alpha}$ is the friction force due to collisions. Using the ideal gas law $P_{\alpha} = n_{\alpha} k_B T$, and assuming quasi-neutrality, $n_{e}(z)=n_{p}(z)=n(z) \quad \forall z$, and the equality of temperatures, $T_{e}(z)=T_{p}(z)=T(z) \quad \forall z$, we subtract the equilibrium equations for each species to isolate the electric field

\begin{equation}\label{Pannekoek+collisions}
    E= E_{PR} -  \frac{1}{2ne} \sum_{\alpha \in \{e,p\}}\mathrm{sign}\left(e_{\alpha}\right)R_{\alpha} + \frac{1}{2ne}\sum_{\alpha \in \{e,p\}}\mathrm{sign}\left(e_{\alpha}\right)\frac{d \left(T_{\alpha} n\right)}{dz}   \quad,
\end{equation}
$E_{PR}$ is the Pannekoek-Rosseland field defined as
\begin{equation}\label{PannekoekRosselandfield}
    E_{PR} = -\frac{\left(m_p-m_e\right)}{2e} g \quad.
\end{equation}
Even if the electron and proton temperatures are equal, the system may not be in thermal equilibrium, and thus $R_{\alpha} \neq 0$ due to collisional effects. Therefore, the total electric field must include an additional ambipolar contribution to restore charge neutrality. Taking the divergence of the electric field yields

\begin{equation}
    \frac{dE}{dz} = \frac{1}{2e}\frac{d}{dz}\left(\frac{1}{n}\sum_{\alpha \in \{e,p\}} \mathrm{sign}\left(e_{\alpha}\right) \left(\frac{d (T_{\alpha} n)}{dz}-R_{\alpha}\right)\right) \quad,
\end{equation}

indicating that a self-consistent, spatially varying electric field must arise from the plasma to sustain quasi-neutrality. This ambipolar electric field becomes essential when collisional terms are included, as the classical Pannekoek-Rosseland field alone is insufficient. Given that \citep{Barbieri2023temperature,Barbieri2024b} utilize the HMF model to approximate electrostatic interactions, a key question arises: can the HMF model, which retains only the first Fourier mode, reproduce the ambipolar effect and maintain charge neutrality in the presence of gravity? If not, can the model be suitably modified to include such effects by extending the Fourier mode expansion?

\begin{figure}
    \centering    \includegraphics[width=0.99\columnwidth]{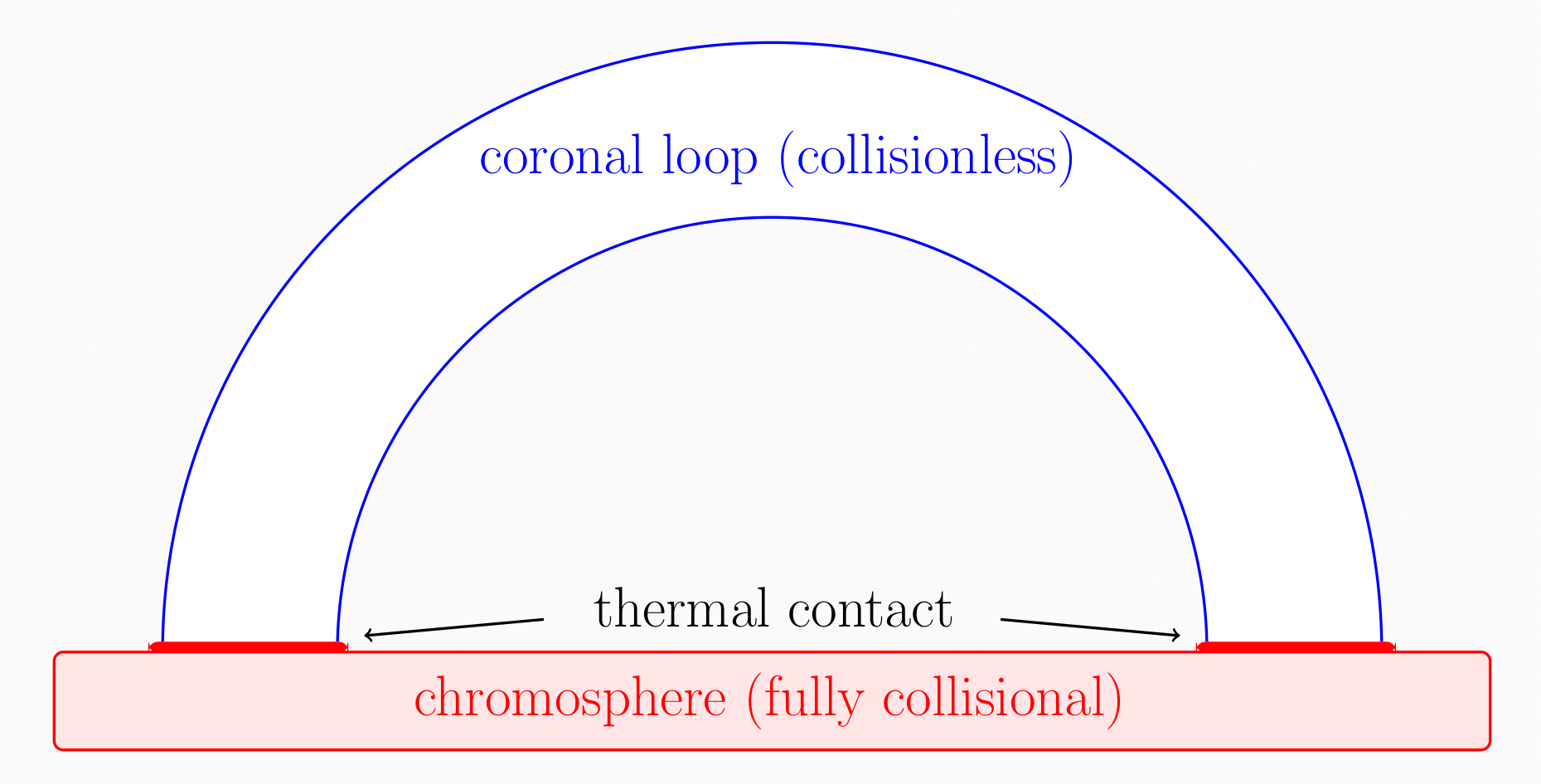}
    \caption{Schematics of the loop model. The coronal plasma in the loop is treated as collisionless and in thermal contact with a fully collisional chromosphere (modelled as a thermal boundary).
    }
    \label{fig:Loopscheme}
\end{figure}
\section{The two-component gravitationally bound plasma model}\label{sec2}

We consider geometrically confined plasma structures, specifically coronal loops, which are prevalent throughout the solar atmosphere (e.g. \cite{2005psci.book.....A}). Each loop is modeled as a semicircular tube of length $2L$  and cross-sectional area $S$ . The plasma inside the loop is treated as a one-dimensional, collisionless, two-species electrostatic system subject to solar gravity. The electrostatic interaction is modeled via a multimode Fourier expansion of the potential, truncated at a chosen mode $N_n$, while enforcing central symmetry at the loop apex. The equations of motion for each particle $j$ are given by

\begin{equation}\label{equationsofmotion}
m_{\alpha} \ddot{x}_{j,\alpha}=eE \left(x_{j,\alpha}\right)+g m_{\alpha}\sin{\left(\frac{\pi x_{j,\alpha}}{2L}\right)}\quad, 
\end{equation}
$x \in [-L,L]$  denotes the curvilinear coordinate along the loop. The self-consistent electric field \footnote{We note that the expression for the electrostatic field expanded in Fourier modes is general and applies both in the presence and absence of collisions. However, given that collisions are not included in the present model, we do not expect to recover Eq. \eqref{Pannekoek+collisions} with all its terms. Furthermore, we point out that, unlike in the previous section, we are now working in the curvilinear coordinates of the loop, which are related to the height $z$ through Eq. $z=\left( 2L/\pi \right)\cos{\left(\pi x/2L\right)}$
} is expressed as
\begin{equation}\label{electricfield}
E(x)=8~\mathrm{sign}{\left(e_{\alpha}\right)}e \cdot n_SN\sum_{n=1}^{N_n} \frac{Q_{n}}{n}\sin{\left(\frac{\pi n x}{L}\right)} \quad.
\end{equation}
$2N$ is the total number of particles of the species $\alpha$, $n_S$ is the surface number density and $Q_n$ are the charge imbalance parameters defined as
\begin{equation}\label{Chargeimbalances}
Q_n=\sum_{\alpha \in \{e,i\}}\mathrm{sign}{\left(e_{\alpha}\right)}q_{n,\alpha}\quad,
\end{equation}
with the stratification parameters $q_{n,\alpha}$  given by
\begin{equation}\label{Stratificationparameters}
    q_{n,\alpha}=\frac{1}{N}\sum_{j=1}^{N} \cos{\left(\frac{\pi nx_{j,\alpha}}{L}\right)}\quad.
\end{equation}
Physically, $q_{n,\alpha} \approx -1$ corresponds to particles concentrated near the base of the loop, $q_{n,\alpha} \approx 0$  to a uniform distribution, and $q_{n,\alpha} \approx 1$ to particles clustered near the loop top. A non-zero $Q_n$ implies a charge imbalance at spatial scale $L/n$. The Fourier components $\rho_n$ of the charge density $\rho(x)$  are related to $Q_n$ via the Poisson equation

\begin{equation}\label{Poissonequation}
    \frac{\partial^2 \phi}{\partial x^2}=-4\pi e\rho \left(x\right) \quad \rightarrow \rho_n = -e n_0 Q_n \quad n_0=\frac{N n_S}{L} \quad,
\end{equation}
indicating that the $Q_n$ directly determine the spatial structure of the charge density at the spatial scale $L/n$.

The loop is assumed to be in ideal thermal contact with a lower boundary representing the fully collisional chromosphere. However, as discussed in the Introduction, the upper chromosphere and the base of the transition region are highly dynamic environments, characterized by intense, short-lived, and small-scale brightenings \citep{Dere:1989ux, Teriaca:2004wy, Peter:2014uz, Tiwari:2019us, Berghmans:2021wl}.

\begin{figure}
\centering
    
 \includegraphics[width=0.99\columnwidth]{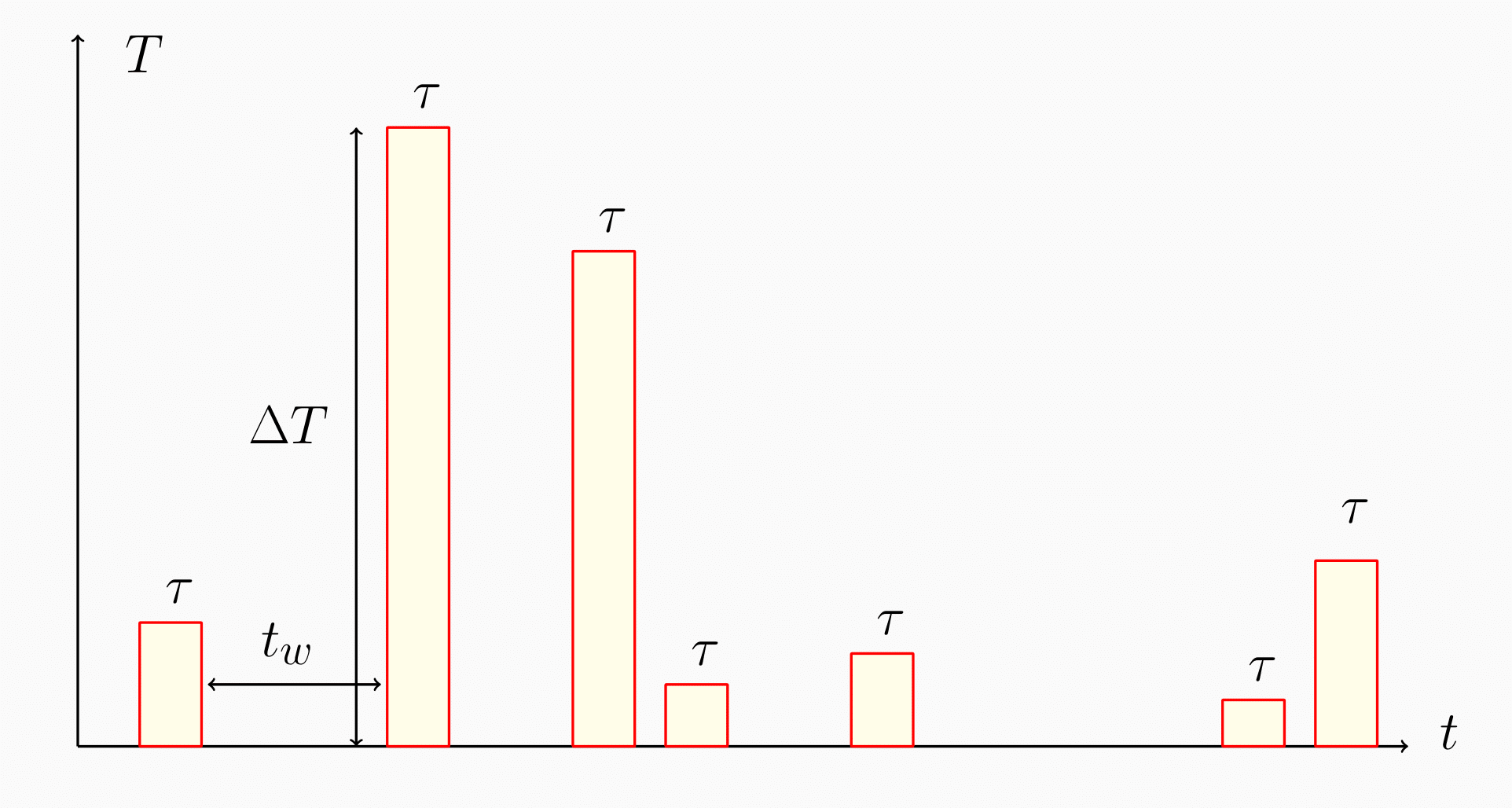}
    \caption{The illustration shows the time evolution of the temperature at the thermal boundary. During intervals of duration $\tau$ , the temperature is increased by an amount $\Delta T$, while during the waiting times $t_w$, it is to the baseline value $T_0$.
    }
    \label{fig:Temperatureflucscheme}
\end{figure}

We model these brightenings as temperature pulses as illustrated in Fig \ref{fig:Temperatureflucscheme}. When the boundary temperature is held constant, the loop plasma reaches thermal equilibrium (isothermal loop). However, the dynamic behaviour of the chromosphere introduces stochastic heating events, which will be analyzed in Section 5.2.

\subsection{The system of units}
We now proceed to express the equations of motion in dimensionless form. We introduce characteristic units
%The following units are defined for velocity, mass and length
\begin{equation}\label{Setsofunits}
\begin{gathered}
    v_0=\sqrt{\frac{k_B T_{0}}{m_e}}, \quad m_0=m_e, \quad  L_0=\frac{L}{\pi} \quad,
\end{gathered}
\end{equation}
and define the dimensionless variables
\begin{equation}
        M=\frac{m_p}{m_e}, \quad
        C=\frac{8 e^2 L^2 n_0}{\pi k_B T_0}, \quad
        \tilde{g}=\frac{g L m_p}{\pi k_B T_{0}}\quad.
\end{equation}
The quantities $C$ and $\tilde{g}$ quantify the strength of the self-electrostatic energy and of the gravitational energy of the electrons in units of thermal energy. 
In these units, denoting with $\theta$ the dimensionless spatial coordinate, the equations of motion become
\begin{equation}\label{HMF2Sdimensionless}
        M_{\alpha} \ddot{\theta}_{j,\alpha}= \mathrm{sign}{\left(e_{\alpha}\right)} E \left(\theta_{j,\alpha}\right)+\tilde{F}\left(\theta_{j,\alpha}\right) \quad,
\end{equation}
the expressions for the external forces and the electrostatic field are
\begin{equation}\label{HMF2Sdimensionless2}
        \tilde{F} \left(\theta_{j,\alpha}\right)= \tilde{g}_{\alpha} \sin{\left(\frac{\theta_{j,\alpha}}{2}\right)}, \quad E(\theta)=  C\sum_{n=1}^{N_n}\frac{Q_n}{n}\sin{\left( n\theta \right)} \quad,
\end{equation}
and
\begin{equation}\label{HMF2Sdimensionless3}
        Q_n=\sum_{\alpha \in \{p,e\}} \mathrm{sign}{\left(e_{\alpha}\right)}q_{n,\alpha},\quad q_{n,\alpha}=\frac{1}{N}\sum_{j=1}^{N}\cos{\left(n\theta_{j,\alpha}\right)} \quad.
\end{equation}
In the aforementioned equations, $M_{\alpha}$ is equal to the mass ratio $M=m_p/m_e$ for protons and $1$ for electrons, while $\tilde{g}_{\alpha}$ is equal to $\tilde{g}$ for protons and $\tilde{g}/M$ for the electrons. Henceforth, all quantities are reported in dimensionless units unless specified otherwise.

\subsection{Vlasov dynamics}\label{subsec:vlasovdynamics}
In the mean-field limit, the dynamics of the phase-space distribution functions is governed by a system of two Vlasov equations
\begin{equation}\label{Vlasovequations}    \frac{\partial f_\alpha}{\partial t}+\frac{p}{M_{\alpha}}\frac{\partial f_{\alpha}}{\partial \theta}+F_{\alpha}[f_{\alpha}]\frac{\partial f_{\alpha}}{\partial p}=0, \quad  F_\alpha=-\frac{\partial H_{\alpha}}{\partial \theta} \quad,
\end{equation}
where $f_\alpha$ are the distribution functions for both species and $H_{\alpha}$ are the mean-field Hamiltonians
\begin{equation}\label{Mean-field-Hamiltonians}
        H_\alpha=\frac{p^2}{2M_\alpha}+ V_{\alpha}, \qquad V_{\alpha} \left( \theta \right)=\mathrm{sign}{\left(e_\alpha \right)} \phi \left(\theta \right)+2\tilde{g}_{\alpha}\cos{\left(\frac{\theta}{2}\right)} \quad,
\end{equation}
$\phi$ is given by
\begin{equation}\label{selfelectrostaticpotential}
   \phi \left(\theta \right)= C \sum_{n=1}^{+\infty} \frac{Q_n}{n^2} \left(\cos{\left(n\theta \right)}+\left(-1\right)^{n+1}\right) \quad, 
\end{equation}
and the $Q_n$ are given by
\begin{equation}\label{qVlasov}
        Q_n[f_{\alpha}]=\sum_{\alpha \in \{e,p\}} \mathrm{sign}{\left(e_{\alpha}\right)} q_{n,\alpha}[f_{\alpha}], \quad
        q_{n,\alpha}[f_{\alpha}]=\int_{-\pi}^{\pi} d\theta \int_{-\infty}^{\infty}dp \cos{\left(n\theta\right)} f_{\alpha}\left(\theta,p\right) \quad.
\end{equation}

\section{Reproducing the Pannekoek-Rosseland field and charge neutrality with the multi-mode model}\label{sec3}
In this section, we demonstrate that the multi-mode electrostatic interaction model is capable of reproducing both the Pannekoek-Rosseland (PR) potential and the charge neutrality condition in the limit of a large number of Fourier modes. According to Jeans’ theorem, all stationary solutions of the Vlasov equation must depend solely on the mean-field Hamiltonians. Consequently, the distribution functions for each species can be expressed as

\begin{equation}\label{Stationarysolution}
    f_{\alpha} \left(\theta,p \right)=f_{\alpha}\left(H_{\alpha}\right) \quad,
\end{equation}
by combining the above equation \eqref{Stationarysolution} with Eq.\eqref{qVlasov} we obtain
\begin{equation}\label{Stationarysystemdistr}
    Q_n=\sum_{\alpha \in \{e,p\}} \mathrm{sign}{\left(e_{\alpha}\right)} \int_{-\infty}^{+\infty} dp \int_{-\pi}^{+\pi} d\theta\cos{\left(n \theta\right)} f_{\alpha}\left(H_{\alpha}\right) \quad.
\end{equation}
Using the kinetic definition of number density, Eq. \eqref{Stationarysystemdistr} can be rewritten as
\begin{equation}\label{Stationarysystemdensity}
    Q_n=\sum_{\alpha \in \{e,p\}} \mathrm{sign}{\left(e_{\alpha}\right)} \int_{-\pi}^{+\pi} d\theta \cos{\left(n\theta \right)} n_{\alpha}\left(V_{\alpha}\right) \quad.
\end{equation}
 
$V_{\alpha}$ is the total potential for each species. Since $V_{\alpha}$ depends on all $Q_n$, this results in a system of coupled non-linear equations. In general, this system cannot be solved analytically for the number densities $n_{\alpha}$, but an analytical solution can be obtained under broad assumptions. First, we assume that the functional form of the number density is identical for both species

\begin{equation}\label{conditiondensities}
    n_{\alpha}\left(V_{\alpha}\right)=n \left(V_{\alpha}\right) \quad.
\end{equation}
This assumption is satisfied by several well-known distribution functions, including thermal distributions, kappa distributions \citep{Scudder1992a,Scudder1992b}, and superstatistics \citep{BECKsuper}, all widely used in theoretical modeling and observations of stellar atmospheres \citep{Dudik_2017,lazar2021kappa,Barbieri2023temperature}.

We now parameterize the Pannekoek-Rosseland potential along the loop's curvilinear coordinate $\theta$ as

\begin{equation}\label{PannekoekRosselandpotential}
    \phi_{PR}\left(\theta \right)= \sum_{\alpha \in \{e,p\}}  \mathrm{sign}\left(e_{\alpha}\right) \frac{\tilde{g}_{\alpha}}{2} \mathrm{cos}\left(\frac{\theta}{2}\right) \quad,
\end{equation}
which can be expanded in a Fourier series as

\begin{equation}
    \phi_{PR}\left(\theta \right)=\sum_{k=0}^{+\infty}c_k \cos{\left(k\theta \right)}, \quad  c_k= \sum_{\alpha \in \{e,p\}} \mathrm{sign}\left(e_{\alpha}\right) \tilde{g}_{\alpha}\frac{\left(-1\right)^k}{\left(1-4k^2\right)} \quad.
\end{equation}
The PR potential can be reproduced by the electrostatic interactions if the following condition holds

\begin{equation}\label{equationQn1}
        C\sum_{n=1}^{N_n}\frac{Q_n}{n^2}\left(\cos{\left(n\theta\right)}+\left(-1\right)^{n+1}\right)=\sum_{k=0}^{+\infty}c_k \cos \left(k\theta \right) \quad.
\end{equation}
This can be rewritten as
\begin{equation}\label{equationQn2}
C\left(\sum_{n=1}^{N_n}\frac{Q_n}{n^2}\cos{\left(n\theta \right)}+Q_0\right) = \sum_{k=0}^{+\infty}c_k \cos\left(\frac{ \pi k x}{L}\right), \qquad Q_0=\frac{\left(-1\right)^n Q_n}{n^2} \quad.
\end{equation}
This condition is satisfied only if the number of Fourier modes $N_n \rightarrow +\infty$, and the $Q_n$ coefficients satisfy

\begin{equation}\label{Qnsolution}
    Q_n = C_q \frac{n^2 \left(-1\right)^n}{\left(1-4n^2\right)},\quad C_q= \frac{1}{C} \sum_{\alpha \in \{e,p\}} \mathrm{sign}\left(e_{\alpha}\right) \tilde{g}_{\alpha} \quad.
\end{equation}
This ensures that the electrostatic potential matches the PR potential exactly. Furthermore, if Eq. \eqref{conditiondensities} is satisfied, then the right-hand side of Eq. \eqref{Stationarysystemdensity} vanishes, implying $Q_n=0 \quad \forall n$. However, Eq. \eqref{Qnsolution} shows that $Q_n \neq 0$, unless $C_q$ is negligible.

For typical coronal plasma conditions (e.g., 
$2L \sim 60 \mathrm{Mm}$, $n_0 \sim 10^{9} \mathrm{cm}^{-3}$, $T_0 = 10^6 K$), we find $C_q \sim 10^{-20}$ , which justifies the approximation $Q_n \sim 0$ , and hence the quasi-neutrality condition is satisfied.
The coefficients $Q_n$
are related to the Fourier modes of the charge density $\rho(x)$ through Eq. \eqref{Poissonequation}. Since $|Q_n| \rightarrow C_q/4$ for large $n$, the Fourier series of $\rho(x)$ diverges unless it is truncated at a finite mode $\hat{n}$, satisfying

\begin{equation}
    \hat{n} Q_{\hat{n}} \sim 1 \quad \mathrm{and} \quad \hat{n} >> 1 \quad.
\end{equation}
leading to
\begin{equation}\label{maximummodes}
   \hat{n} \sim 4/C_{q} \quad \forall n< \hat{n} \quad.
\end{equation}
%Now, since $C_q \sim 10^{-20}$ means that up to $\hat{n} \sim 10^{20}$ Fourier modes, Eq. \eqref{Qnsolution} solve the non-linear system given by Eq. \eqref{Stationarysystemdensity}. Since $Q_n \sim 0 \quad \forall n < \hat{n}$ for Eqs. \eqref{Poissonequation}, the condition $\rho_n \sim 0 \quad \forall n < \hat{n}$ is also satisfied and thus the quasi-neutrality.

Now, since $C_q \sim 10^{-20}$, for equation \eqref{maximummodes} we have that $\hat{n} \sim 4 \cdot 10^{20}$. This implies that the solution of the system of equations \eqref{Qnsolution} is given by Eq.\eqref{Stationarysystemdensity} up to $n<\hat{n}$. Now, since $Q_n \sim 0$ for all $n< \hat{n}$ , and given that the charge density components $\rho_n$ are proportional to  $Q_n$ via Eq.  \eqref{Poissonequation}, it follows that  $\rho_n \sim 0$ for all $n<\hat{n}$ as well.

In conclusion, we have shown that electrostatic interactions can reproduce the Pannekoek–Rosseland potential and enforce quasi-neutrality down to a very small spatial scale proportional to $L/\hat{n}$, in the realistic limit where electrostatic forces dominate over gravity, i.e., $C_q <<1$. This behaviour appears to be general and not limited to loop geometry, as further demonstrated in Appendix A for a plane-parallel atmosphere. Now a natural question arises: where does this electric field come from?

\begin{figure}
    \centering
    \includegraphics[width=0.99\columnwidth]{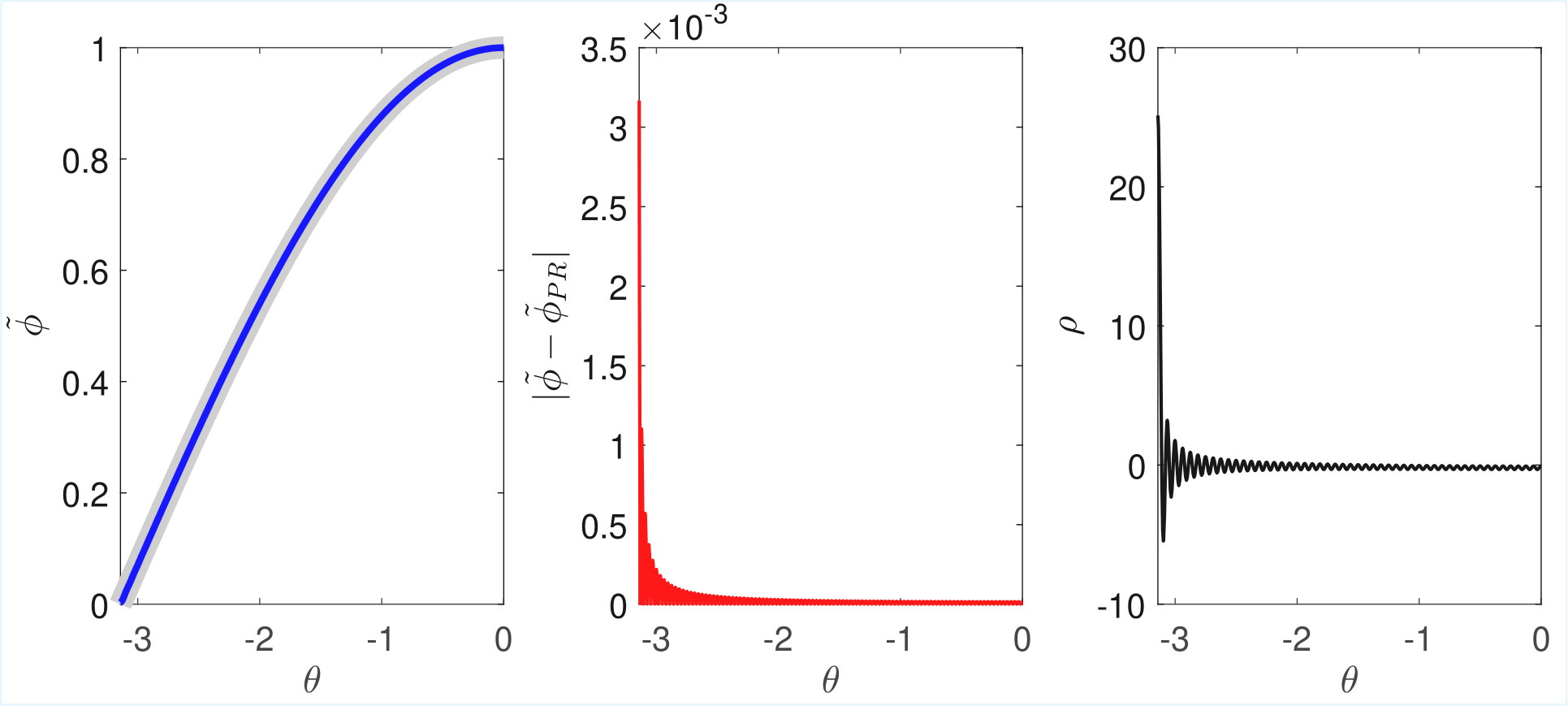}
    \caption{Left panel: the self-consistent electrostatic potential $\phi$ (blue) and the Pannekoek-Rosseland potential $\phi_{PR}$ (gray), plotted as functions of the curvilinear coordinate $\theta$, both normalized by the factor $\sum_{\alpha \in \{e,p\}} \mathrm{sign} \left(e_{\alpha}\right) \tilde{g}_{\alpha}$. The resulting rescaled potentials are denoted as $\tilde{\phi}$ and $\tilde{\phi}_{PR}$ , respectively. The self-consistent potential is computed from Eq.\eqref{electrostaticpotentialtruncated}, while the Pannekoek-Rosseland potential is obtained from Eq.\eqref{PannekoekRosselandpotential}.
    Center panel: the absolute difference $|\tilde{\phi}-\tilde{\phi}_{PR}|$ as a function of $\theta$, highlighting the spatial deviation between the two potentials.
    Right panel: the charge density $\rho(\theta)$ plotted as a function of $\theta$, calculated using Eq.~\eqref{Chargedensityruncated}.
    }
    \label{Panneanddensity}
\end{figure}

To investigate where any residual charge imbalance occurs, we consider the case $C_q \sim 400$, implying $\hat{n}=100$ according to \eqref{maximummodes}. The electrostatic potential is then approximated by
\begin{equation}\label{electrostaticpotentialtruncated}
    \phi \left(\theta \right) \cong \sum_{k=0}^{\hat{n}} c_{k}\cos{\left( k\theta \right)} \quad,
\end{equation}
and the total charge density becomes
\begin{equation}\label{Chargedensityruncated}
    \rho \left(\theta \right)\cong -\sum_{n=1}^{\hat{n}}Q_n \cos{\left(n\theta \right)} \quad.
\end{equation}
Figure \ref{Panneanddensity} illustrates that the self-consistent electrostatic potential (blue) closely approximates the PR potential (gray) throughout the loop. The middle panel shows that the discrepancies are more pronounced near the base of the loop. These differences are associated with a localized charge imbalance, as depicted in the right panel. This analysis indicates that the PR potential primarily arises from a thin charge layer at the base of the loop, while the upper regions remain approximately quasi-neutral. This outcome is anticipated, as the gravitational force attains its maximum at the base, whereas the electrostatic field vanishes at that point.

\begin{figure}
    \centering
    \includegraphics[width=0.99\columnwidth]{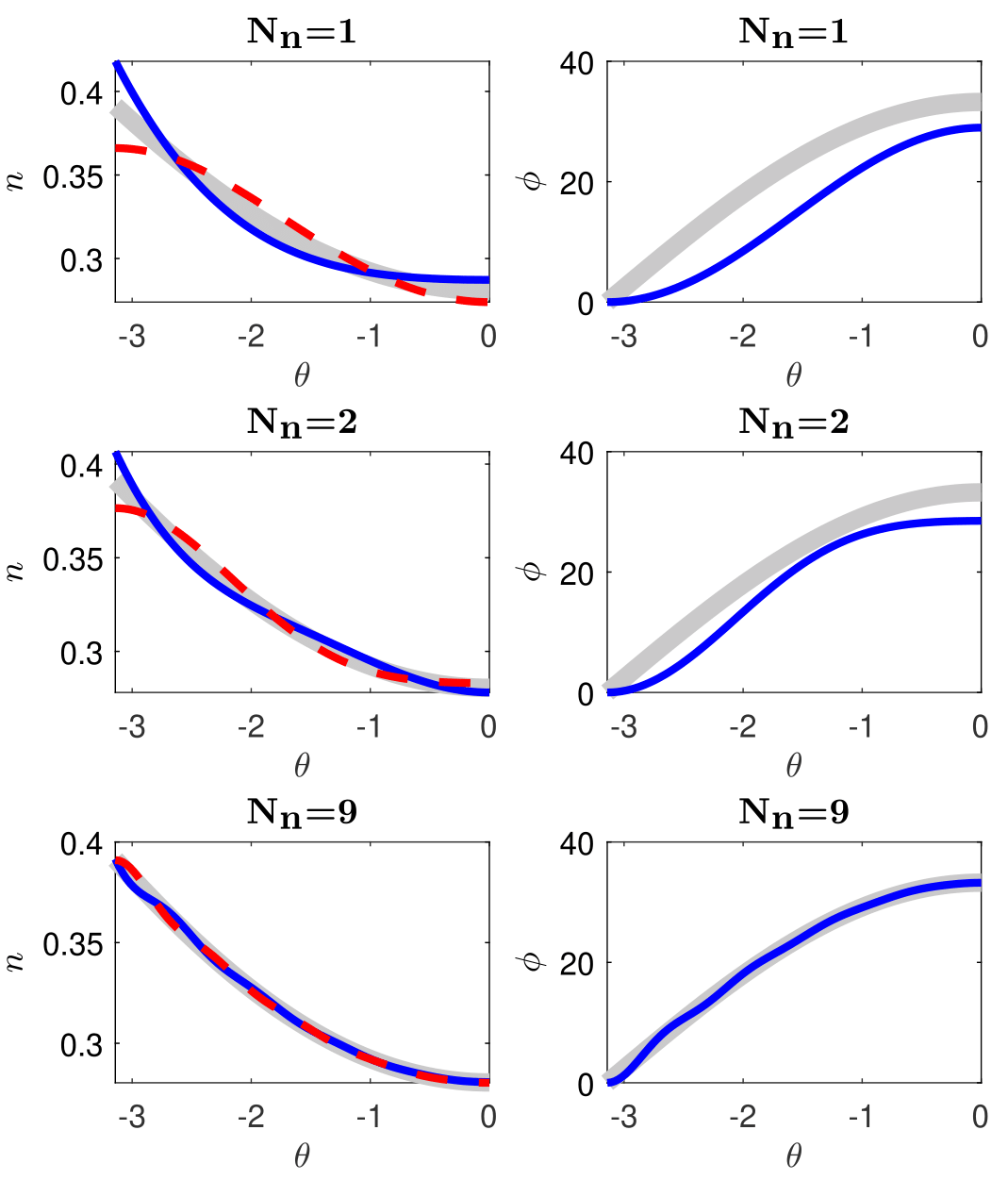}
    \caption{Top row, left panel: electron (red) and proton (blue) number densities as functions of the curvilinear coordinate along the loop. The densities are computed using Eq.\eqref{densitythermal} with a single Fourier mode ($N_n=1$), as indicated in the subplot title. The grey curve shows the reference density profile corresponding to the Pannekoek-Rosseland potential. Top row, right panel: the Pannekoek-Rosseland potential (grey), calculated using Eq.\eqref{PannekoekRosselandfield}, and the self-consistent electrostatic potential (blue), computed using Eq.~\eqref{selfelectrostaticpotential} with one Fourier mode. Middle and bottom rows: same as the top row, but for $N_n=2$ and $N_n=9$, respectively.
    }
    \label{DensityandPannemulti}
\end{figure}

\section{Validation of the multi-mode electrostatic model}\label{sec4}
In the previous section, we demonstrated that under the following assumptions
\begin{itemize}
    \item The number density satisfies $n_{\alpha} \left(V_{\alpha}\right)=n \left(V_{\alpha}\right)$.
    \item The parameter $C_q << 1$, that is, electrostatic interactions are much stronger than gravitational forces,
\end{itemize}
the multi-mode electrostatic model increasingly approximates charge neutrality and the Pannekoek-Rosseland potential as the number of Fourier modes increases. We now validate this conclusion for different types of distribution functions.

\subsection{Thermal solution for multi-modes models}\label{subsec:thermal}

Among all the stationary solutions of the Vlasov equation, the thermal solution is given by
\begin{equation}\label{thermalsolution}
    f_{\alpha}\left(H_{\alpha}\right)=\frac{1}{Z_{\alpha}}e^{-\frac{H_{\alpha}}{T}} \qquad \mathrm{and} \qquad Z_{\alpha}=\int_{-\infty}^{+\infty}dp \int_{-\pi}^{\pi}d\theta e^{-\frac{H_{\alpha}}{T}} \quad,
\end{equation}
where $Z_{\alpha}$ is the partition function.

Using the standard kinetic definitions, the number density for each species becomes
\begin{equation}\label{densitythermal}
    n_{\alpha}\left(\theta \right)= n \left(V_{\alpha}\right)=\frac{e^{-\frac{V_{\alpha}\left(\theta \right)}{T}}}{\int_{-\pi}^{\pi}d\theta e^{-\frac{V_{\alpha}\left(\theta \right)}{T}}} \quad . 
\end{equation}
From this expression, it is evident that thermal distributions satisfy the density condition in Eq. \eqref{conditiondensities}. Substituting Eq. \eqref{densitythermal} into Eq. \eqref{Stationarysystemdensity} and performing some algebra yields
\begin{equation}\label{Qnthermal}
    Q_n= \sum_{\alpha \in \{e,p\}}\mathrm{sign}(e_{\alpha}) \frac{\int_{-\pi}^{0} d\theta\cos{\left(n\theta \right)}e^{-\frac{V_{\alpha}\left(\theta \right)}{T}}}{\int_{-\pi}^{0}d\theta e^{-\frac{V_{\alpha}\left(\theta \right)}{T}}}\quad.
\end{equation}
Solving this non-linear system provides the values of $Q_n \quad \forall n$. To ensure the recovery of charge neutrality and the PR potential, the condition $C_q <<1$ must be satisfied.

Figure \ref{DensityandPannemulti} presents the numerical results obtained with dimensionless parameters satisfying $C_q <<1$, specifically: 
$\tilde{g}=32, M=m_p/m_e, C=10^4, T=90$. The left panels show the electron (red) and proton (blue) number density profiles computed via Eq. \eqref{densitythermal}, where the coefficients $Q_n$ are derived by solving Eq. \eqref{Qnthermal} using $N_n=1$ (top row), $N_n=2$ (middle row), and $N_n=9$ (bottom row) Fourier modes. The gray curves represent the density profiles computed using the PR potential from Eq. \eqref{PannekoekRosselandpotential}. As the number of Fourier modes increases, the species densities converge towards the PR profile, indicating improved charge neutrality. The right panels of Figure \ref{DensityandPannemulti} show the corresponding electrostatic potential computed from Eq. \eqref{selfelectrostaticpotential} (blue), alongside the PR potential (gray). The convergence of the self-consistent potential towards the PR solution with increasing $N_n$ confirms the analytical predictions of Section 4.

\subsubsection{Validation via numerical simulations}
We now demonstrate that the dynamic model introduced in Section 3 relaxes towards the thermal equilibrium configuration described above, provided that the thermal boundary is held at a constant temperature $T$. We initialize the system with isothermal distributions for both electrons and protons, accounting for their respective gravitational stratifications

\begin{equation}\label{distrthermal}
    f_{\alpha}(\theta,p)= \frac{e^{-\frac{1}{T}\left(\frac{p^2}{2M_{\alpha}}+2 \tilde{g}_{\alpha} \cos{\left(\frac{\theta}{2}\right)}\right)}}{\sqrt{2\pi M_{\alpha}T}\int_{-\pi}^{\pi}d\theta e^{-\frac{2 \tilde{g}_{\alpha}}{T} \cos{\left(\frac{\theta}{2}\right)}}} \quad.
\end{equation}
This configuration is not dynamically stable, as it yields distinct stratifications for the two species. As a result, an ambipolar electric field emerges and progressively acts to equalize the species densities, enforcing charge neutrality. This effect can only be reproduced if a sufficient number of Fourier modes is included in the electrostatic model.

\begin{figure}
    \centering
    \includegraphics[width=0.99\columnwidth]{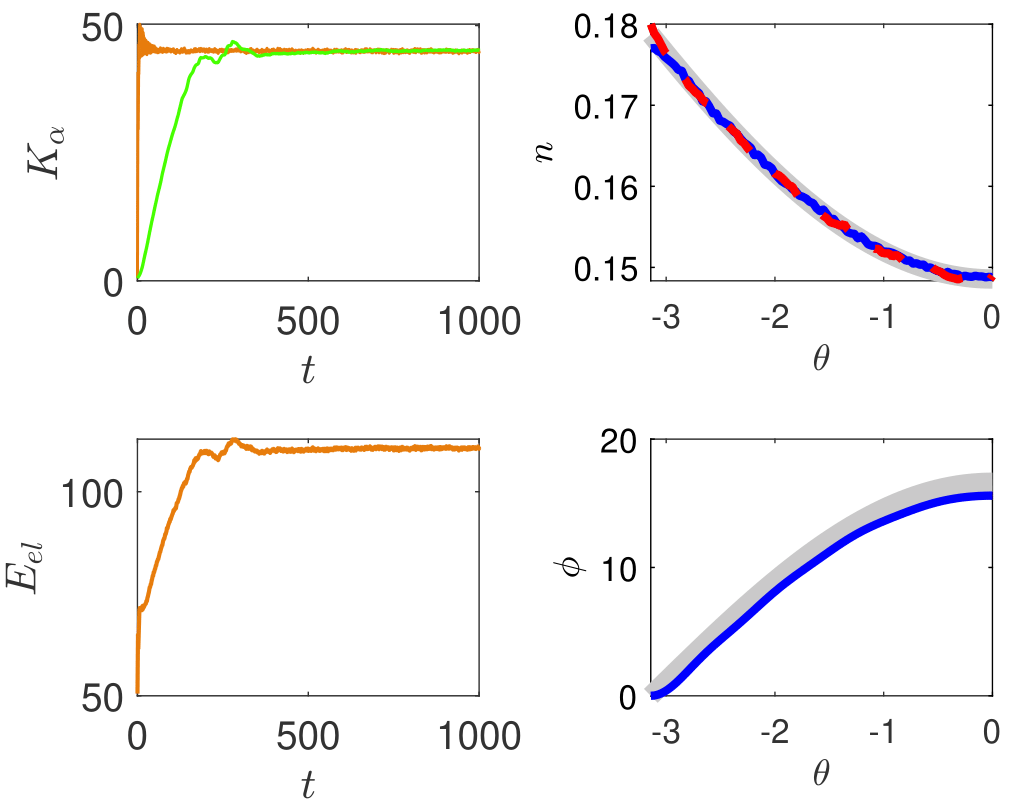}
    \caption{Top left panel: time evolution of the kinetic energies $K_{\alpha}$ of protons (green curve) and electrons (orange curve). Top right panel: number densities of electrons (blue) and protons (red). The grey curves correspond to the theoretical density profile obtained from Eq.\eqref{densitythermal}, where the electrostatic potential $\phi$ is replaced by the Pannekoek potential given in Eq.\eqref{PannekoekRosselandpotential}. Bottom left panel: time evolution of the total electrostatic energy $E_{el}$, evaluated numerically via Eq.\eqref{electrostaticenergy}.
    Bottom right panel: self-consistent electrostatic potential $\phi$, computed from simulations using Eq.\eqref{selfelectrostaticpotential}. The grey curve represents the Pannekoek potential given by \eqref{PannekoekRosselandpotential}.
    }
    \label{Thermaldynamics}
\end{figure}

We integrate numerically Eq. \eqref{HMF2Sdimensionless} \footnote{The details of the numerical approach are explained in \citep{Barbieri2024b}.} using $N_n=9$ modes and the same parameter values as in Fig. \ref{DensityandPannemulti}, for $N=2^{17}$ particles. The top-left panel of Fig. \ref{Thermaldynamics} shows the time evolution of the kinetic energies of electrons (orange) and protons (green), defined as
\begin{equation}\label{stratpar1}
    K_{\alpha} = \frac{1}{N}\sum_{j=1}^{N} \frac{p_{j,\alpha}^2}{2M_{\alpha}} \quad .
\end{equation}
Both kinetic energies relax to an asymptotic value, consistent with thermal equilibrium. In this state, the electron and proton densities coincide and match the PR density profile (gray), shown in the top-right panel. The bottom-left panel illustrates the time evolution of the total electrostatic energy
\begin{equation}\label{electrostaticenergy}
    E_{el}= \frac{C}{2} \sum_{n=1}^{N_n} \frac{Q_n^2}{n^2} \quad.
\end{equation}
This quantity also relaxes to an asymptotic value. The electrostatic potential corresponding to this asymptotic state (blue), shown in the bottom-right panel, closely matches the PR potential (gray), confirming the model’s ability to self-consistently reproduce the equilibrium solution.

\subsection{The multitemperature solution for multi-modes models}\label{subsec:multithermal}
\begin{figure}
    \centering
    \includegraphics[width=0.99\columnwidth]{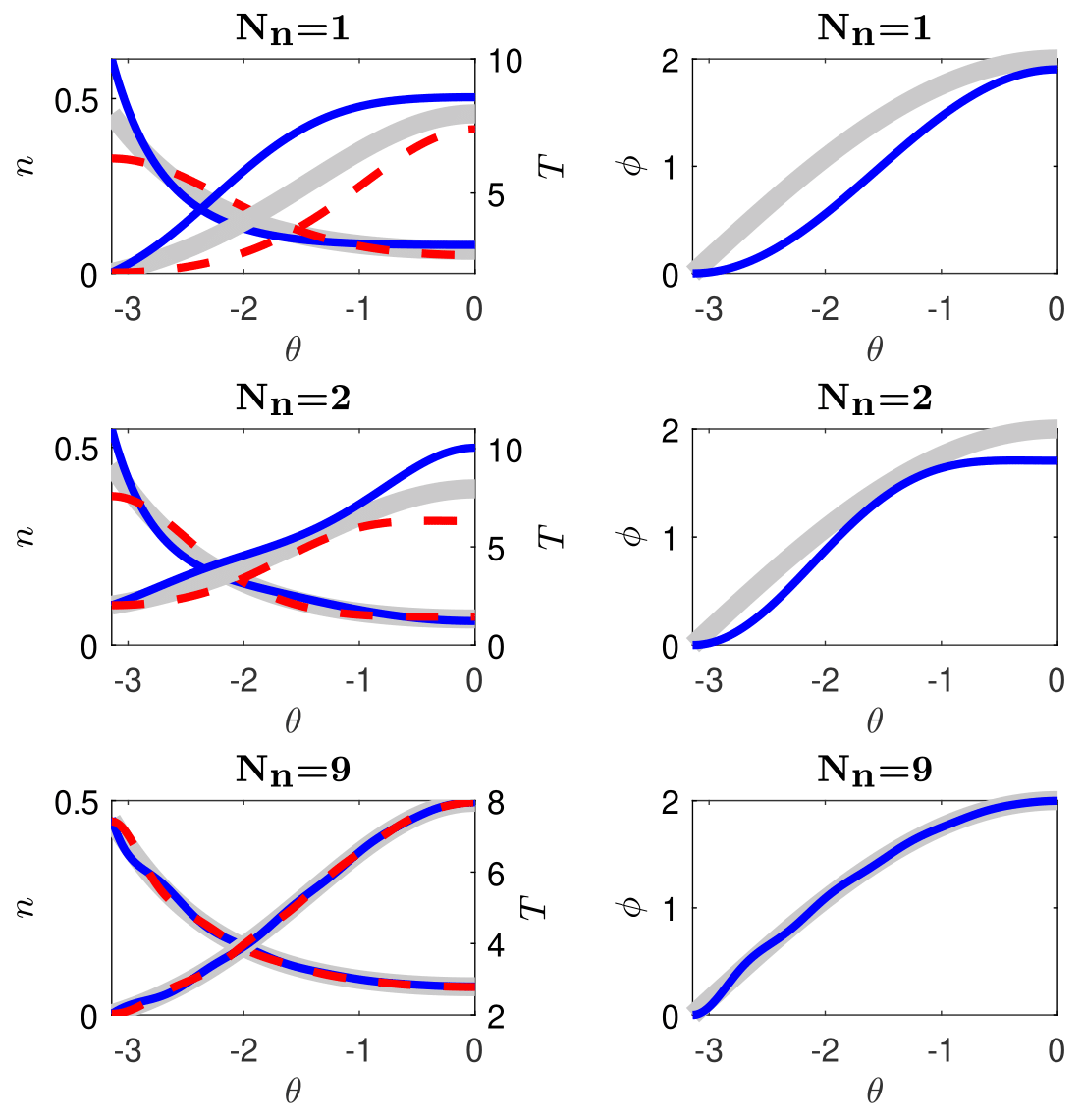}
    \caption{Top row, left panel: electron (red) and proton (blue) number densities (globally decreasing functions) and temperatures (globally increasing functions) as functions of the curvilinear coordinate along the loop. The densities and temperature are computed using Eq.\eqref{densitythermal} with a single Fourier mode ($N_n=1$), as indicated in the subplot title. The grey curve shows the reference temperature and density profiles corresponding to the Pannekoek-Rosseland potential. Top row, right panel: the Pannekoek-Rosseland potential (grey), calculated using Eq.\eqref{PannekoekRosselandfield}, and the self-consistent electrostatic potential (blue), computed using Eq.~\eqref{selfelectrostaticpotential} with one Fourier mode. Middle and bottom rows: same as the top row, but for $N_n=2$ and $N_n=9$, respectively.
    }
    \label{DensityandPannebitemp}
\end{figure}

The thermal solution presented above is valid under the assumption of a static thermal boundary at fixed temperature $T$. However, as discussed in Section 3  as well as in the introduction, our primary interest lies in the case where the thermal boundary exhibits time-dependent temperature fluctuations. As shown in \cite{Barbieri2024b}, if the time scales of these fluctuations are shorter than the relaxation time of the loop at each corresponding temperature, then the system relaxes toward a non-equilibrium stationary state. In this regime, the distribution functions of the plasma particles are described by the following analytical expression

\begin{equation}\label{multitemperature}
    f_{\alpha}\left(\theta,p\right)=\mathcal{N}_{\alpha}\left(A\int_1^{+\infty}dT\frac{\gamma \left(T\right)}{T}e^{-\frac{H_{\alpha}}{T}}+\left(1-A\right)e^{-H_{\alpha}}\right), \quad A=\frac{\tau}{\tau + \langle t_w \rangle_{\eta}} \quad,
\end{equation}
where $\mathcal{N}_{\alpha}$ is the normalization constant ensuring that $f_{\alpha}$ integrates to unity, $\gamma(T)$ is the probability distribution of the temperature increments, $\tau$ is the duration of the heating events, $\langle t_w \rangle$ the average waiting time between events over the probability distribution $\eta(t_w)$. Using the standard kinetic definitions, we compute the number density and kinetic temperature for each species

\begin{equation}\label{NEqdensitywithC}
   n_{\alpha}\left(\theta \right)=n \left(V_{\alpha} \right)=\frac{A\int_{1}^{+\infty}dT\frac{\gamma \left(T\right)}{\sqrt{T}}e^{-\frac{V_{\alpha}}{T}}+\left(1-A\right)e^{-V_{\alpha}}}{A\int_{1}^{+\infty}dT\frac{\gamma \left(T\right)}{\sqrt{T}}\int_{-\pi}^{\pi} d\theta e^{-\frac{V_{\alpha}}{T}}+\left(1-A\right)\int_{-\pi}^{\pi} d\theta e^{-V_{\alpha}}} \quad,
\end{equation}
\begin{equation}\label{NEqtemperaturewithC}
    T_{\alpha}\left(\theta \right)=T \left(V_{\alpha} \right)=\frac{A\int_{1}^{+\infty}dT\gamma \left(T\right)\sqrt{T}e^{-\frac{V_{\alpha}}{T}}+\left(1-A\right)e^{-V_{\alpha}}}{A\int_{1}^{+\infty}dT\frac{\gamma \left(T\right)}{\sqrt{T}}e^{-\frac{V_{\alpha}}{T}}+\left(1-A\right)e^{-V_{\alpha}}} \quad.
\end{equation}
These multi-temperature distributions also satisfy the condition given by Eq. \eqref{conditiondensities}, required for the system to approach charge neutrality.
By inserting Eq. \eqref{NEqdensitywithC} into the system of equations \eqref{Stationarysystemdensity}, and after some manipulation, we obtain

\begin{equation}\label{Qnmultitemp}
   Q_n=\sum_{\alpha \in \{e,p\}} \mathrm{sign}\left(e_{\alpha} \right)\frac{A\int_{1}^{+\infty}dT\frac{\gamma \left(T \right)}{\sqrt{T}}\int_{-\pi}^{0} d\theta \cos{\left(n\theta \right)}e^{-\frac{V_{\alpha}}{T}}+\left(1-A\right)\int_{-\pi}^{0} d\theta \cos{\left(n\theta \right)}e^{-V_{\alpha}}}{A\int_{1}^{+\infty}dT\frac{\gamma \left(T\right)}{\sqrt{T}}\int_{-\pi}^{0} d\theta e^{-\frac{V_{\alpha}}{T}}+\left(1-A\right)\int_{-\pi}^{0} d\theta e^{-V_{\alpha}}} \quad.
\end{equation}
Solving this system yields the values of $Q_n$ for all modes. As with the thermal case, we demonstrate this behaviour using dimensionless parameters that satisfy the condition $C_q <<1$.
The results are shown in Fig. \ref{DensityandPannebitemp}, where we assume a delta-distribution for the temperature increments

\begin{equation}
    \gamma \left(T\right) = \delta \left(T-\left(1+\Delta T\right)\right) \quad.
\end{equation}
corresponding to a boundary temperature that alternates between two discrete values. The chosen parameters are: $A=0.1,\tilde{g}=2, M=m_p/m_e, C=10^4, \Delta T=90$. In Figure \ref{DensityandPannebitemp}, the left panels display the temperature and number density profiles of electrons (blue) and protons (red dashed), computed from Eqs. \eqref{NEqdensitywithC} and \eqref{NEqtemperaturewithC}, respectively, using $N_n=1,2,9$. The gray lines represent the profiles obtained using the PR potential. As in the thermal case, increasing the number of Fourier modes leads to convergence of the electron and proton profiles toward the PR solutions, indicating that charge neutrality and the correct electrostatic potential structure are recovered. The right panels show the corresponding electrostatic potentials computed using Eq. \eqref{selfelectrostaticpotential} (blue), compared to the PR potential (gray). Once again, convergence is observed as the number of modes increases. In appendix \ref{sec:appendixB} we show a case in which $\gamma(T)$ is not a simple delta-distribution proving the robustness of the present result.

\subsubsection{Validation against numerical simulations}
\begin{figure}
    \centering
    \includegraphics[width=0.99\columnwidth]{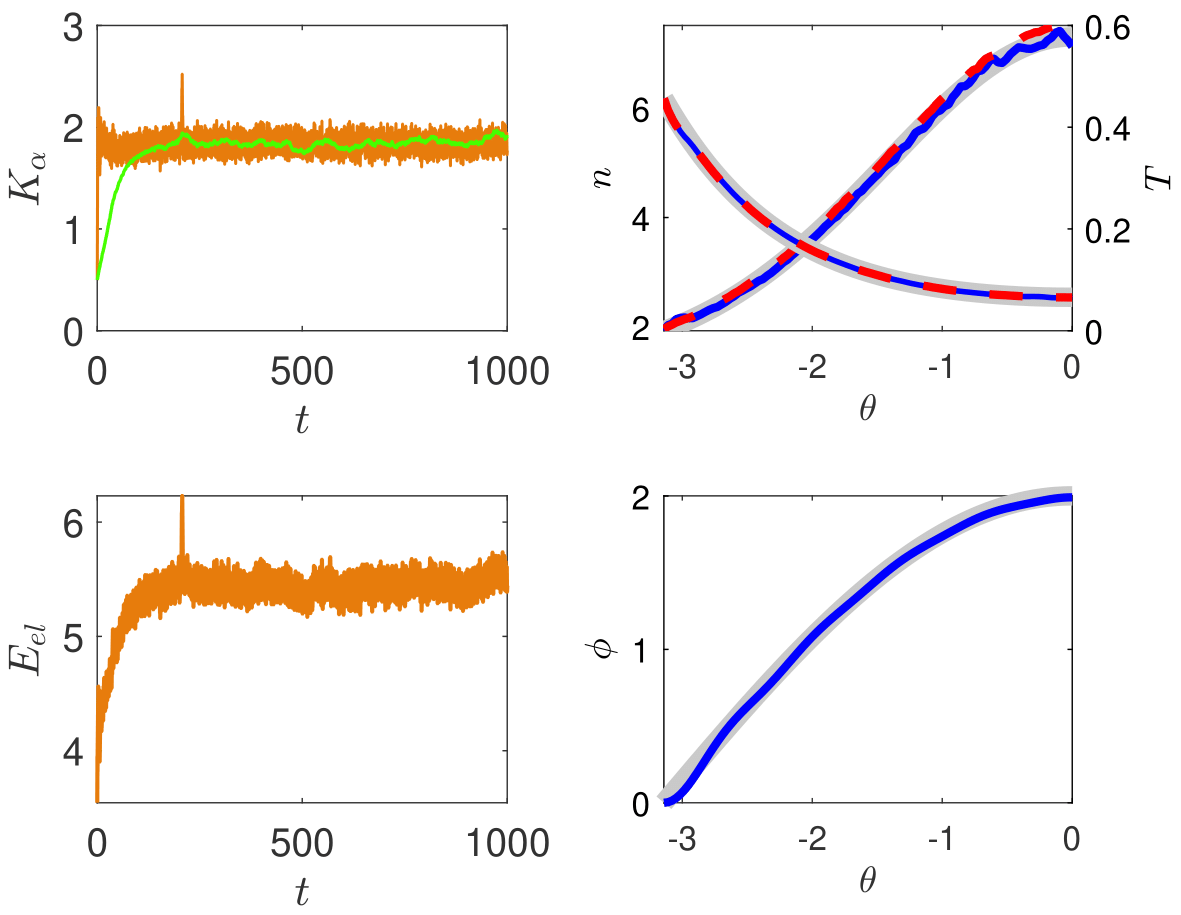}
    \caption{Same quantities and color scheme as in Fig.\ \ref{Thermaldynamics}, but in the multitemperature case regime described in Sec.\ \ref{subsec:multithermal}.
    }
    \label{Multitempdynamics}
\end{figure}

The dynamics governed by the equations of motion \eqref{equationsofmotion}, coupled with the fluctuating thermal boundary, have been numerically simulated following the methodology described in \citep{Barbieri2023temperature,Barbieri2024b}, with the notable difference that the electric field is computed using Eq. \eqref{electricfield}. In this section, we illustrate how the fluctuating thermal boundary drives the overlying collisionless plasma toward a stationary state characterized by a temperature inversion. To this end, the system is initialized in a thermal equilibrium configuration, in which the distribution functions for both species are given by

\begin{equation}\label{densitythermalpanne}
    f_{\alpha}\left(\theta,p\right)= \frac{e^{-\frac{p^2}{2M_{\alpha}}-2 \tilde{g} \cos{\left(\frac{\theta}{2}\right)}}}{\sqrt{2\pi M_{\alpha}}\int_{-\pi}^{\pi}d\theta e^{-2 \tilde{g} \cos{\left(\frac{\theta}{2}\right)}}} \quad \forall \alpha \quad.
\end{equation}
This configuration is not stationary under the influence of the dynamically fluctuating thermal boundary at the base. However, as discussed in Section \ref{sec2}, if a sufficiently large number of Fourier modes is retained in the modeling of the self-consistent electrostatic interactions, we expect the system to recover both charge neutrality and the Pannekoek-Rosseland potential in the stationary state. We report here the results of a simulation conducted with $N_n=9$ modes and using the same set of numerical parameters employed for the production of Fig.\ref{DensityandPannebitemp}. The total number of particles is set to $N=2^{17}$. In analogy with Fig.\ref{Thermaldynamics}, the upper-left panel of Fig.\ref{Multitempdynamics} shows the time evolution of the total kinetic energy of each species $\alpha$, computed according to Eq.\eqref{stratpar1}. In this case as well, the kinetic energies of both species relax toward an asymptotic value. The upper-right panel displays the final stationary profiles of temperature and density computed via the numerical simulation together with their theoretical counterpart, which clearly exhibit a temperature inversion. The theoretical profiles, shown in grey, are computed using Eqs.\eqref{NEqdensitywithC} and \eqref{NEqtemperaturewithC}, where the electrostatic potential $\phi(\theta)$ is replaced by the Pannekoek-Rosseland expression given in Eq.\eqref{PannekoekRosselandpotential}. The lower-left panel illustrates the time evolution of the total electrostatic energy, evaluated using Eq.\eqref{electrostaticenergy}. As observed in the case discussed in Section \ref{subsec:thermal}, this quantity also relaxes to an asymptotic value. Finally, the lower-right panel confirms that the self-consistent electrostatic potential generated by the plasma coincides with the Pannekoek-Rosseland potential at equilibrium.

\section{Summary and perspectives}

The main plasma physics result of the present work is the identification of a self-consistent mechanism to reproduce the ambipolar effect that gives rise to the Pannekoek–Rosseland electric field and ensures charge neutrality in a gravitationally stratified, collisionless plasma. This is achieved by extending the electrostatic interaction model from a single Fourier mode (as in the standard HMF approximation) to multiple modes. Under very general assumptions on the form of the distribution functions, assumptions that are satisfied by thermal, kappa, and superstatistical distributions, we demonstrated that in the limit of many modes, the self-consistent electrostatic potential converges to the classical Pannekoek-Rosseland potential. At the same time, the system exhibits quasi-neutrality apart from a thin layer of charge at the base of the atmosphere which originates the Pannekeok-Rosseland potential. These theoretical predictions were confirmed numerically. We have shown that, under suitable conditions, the distribution functions of both species relax toward a stationary configuration in which charge neutrality and the Pannekoek-Rosseland potential are simultaneously recovered. This was observed both in the thermal equilibrium case (where the boundary temperature is fixed) and in the more physically relevant multitemperature case, where the thermal boundary undergoes random energy fluctuations.

While the results of this work do not aim to provide a general theory of ambipolar electric fields nor a complete description of coronal heating, they represent a foundational step in that direction. Now that we have established how to incorporate the ambipolar effect in a self-consistent way into the model originally introduced in \citep{Barbieri2023temperature,Barbieri2024b}, we are in a position to extend it. In particular, we can include additional ingredients such as collisions, ionization processes, and asymmetric temperature fluctuations at the loop footpoints. These effects will allow us to determine how the ambipolar electric field predicted by the multimode Fourier approach is modified, and how the picture of coronal heating evolves in this more complete framework.

As discussed in Section 2, we expect collisions to modify the self-generated electric field with respect to the classical Pannekoek–Rosseland expression. For example, we aim to investigate the following questions: How many Fourier modes are required to reproduce the modified ambipolar field in the collisional case? In which regions of the atmosphere are the deviations from the Pannekoek–Rosseland field most significant? And how does this modified electric field acts on the temperature inversion produced by gravitational filtering in the collisionless limit? Finally, by including collisions, we can study the interplay between the collisionless gravitational filtering, which produces the temperature inversion effect, and the collisional thermal conduction, which allows heat to flow from the corona into the chromosphere.

Another important extension of the present model concerns the role of partial ionization, especially in the lower part of the transition region, where the ionization process is known to occur. In this regime, deviations from the classical Pannekoek–Rosseland field may become more significant, as the plasma is not yet fully ionized. As altitude increases and ionization progresses, the assumptions of a fully ionized and collisionless plasma, underlying the present treatment, become more accurate. Understanding how the ambipolar electric field evolves across this ionization gradient represents an important development of our model.

We note that in our model we have the central symmetry with respect to the loop apex. This will ensure a zero net flux in the system. By introducing asymmetric temperature fluctuations at the two footpoints. A net flux across the loop appears.

However, the resulting asymmetry does not alter the physical mechanism responsible for the temperature inversion. The inversion still arises because "hot" particles, generated by temperature increments, can climb higher in the gravitational potential well, while "cold" particles remain concentrated near the base and the temperature profile along the loop is no longer symmetric. Although this adds complexity to the model, it does not change the physical origin of the temperature inversion. It represents an interesting extension that we plan to explore in a future work.

We note that generating different temperature increments at the two footpoints of the loop produces a flow into the system, altering the ambipolar electric field, which would no longer be the classical Pannekoek–Rosseland. Another way to obtain a different ambipolar field could be through the introduction of distinct temperature fluctuations for the two species. This leads to different temperature profiles for electrons and protons, and consequently, as evident from Eq. \eqref{Pannekoek+collisions}, an ambipolar electric field that deviates from the classical Pannekoek–Rosseland expression.

Finally, as an additional possible follow-up, it would be interesting to include alpha particles in the modeling, as they contribute significantly to the mass density.

\appendix

\section{Plane-parallel plasma atmosphere}\label{sec:appendixA}

In this appendix, we adopt the CGS system of units. We consider a model of a two-component, plane-parallel, collisionless plasma atmosphere. Particles are subject to both the gravitational field of the central star and to self-consistent electrostatic interactions. Under these assumptions, the distribution functions $f_{\alpha}$ evolve according to the Vlasov equations

\begin{equation}\label{VlasovequationsPP}
    \frac{\partial f_{\alpha}}{\partial t}+\frac{p}{m_{\alpha}}\frac{\partial f_{\alpha}}{\partial z}+F_{\alpha}[f_{\alpha}]\frac{\partial f_{\alpha}}{\partial p} = 0 \quad, F_{\alpha} = -\frac{\partial H_{\alpha}}{\partial z}, \quad H_\alpha=\frac{p^2}{2m_\alpha}+V_{\alpha} \quad,
\end{equation}
the total potential is defined as
\begin{equation}\label{Mean-field-HamiltoniansPP}
        V_{\alpha}(z)=\mathrm{sign}(e_\alpha)\phi(z)+m_{\alpha}g z \quad,
\end{equation}
with $\phi(z)$ being the electrostatic potential, which satisfies the Poisson equation

\begin{equation}\label{PoissonequationPP}
    \frac{d^2 \phi}{d z^2}=-4\pi e\rho(z) \quad.
\end{equation}
Using the Green’s function formalism, the electrostatic potential can be expressed as

\begin{equation}
    \phi(z)=-2\pi e^2 n_0 \sum_{\alpha \in \{e,p\}} \int_{0}^{L} d \tilde{z} n_{\alpha}(z)|z-\tilde{z}| \quad,
\end{equation}
$L$ is the extent of the atmosphere. Assuming that the distribution functions depend only on the Hamiltonians (Jeans theorem), and making the same assumptions as in Eq. \eqref{conditiondensities}, we obtain

\begin{equation}
    \phi(z)=-2\pi e^2 n_0 \sum_{\alpha \in \{e,p\}} \int_{0}^{L} d \tilde{z} n(V_{\alpha})|z-\tilde{z}| \quad.
\end{equation}
Introducing the dimensionless coordinate 
$y = z / L$, we rewrite the expression as

\begin{equation}\label{integralpoissonPP}
    \frac{\phi(z)}{2\pi e^2 n_0 L^2} = \sum_{\alpha \in \{e,p\}} \int_{0}^{1} d \tilde{y} n(V_{\alpha})|y-\tilde{y}| \quad.
\end{equation}
We parameterize the Pannekoek-Rosseland field in this geometry

\begin{equation}\label{PannekoekRosselandpotentialA}
    \phi_{PR}(z)= \sum_{\alpha \in \{e,p\}} \mathrm{sign}(e_{\alpha}) m_{\alpha}gz \quad.
\end{equation}
Substituting $\phi(z)=\phi_{PR}(z)$ into the integral equation above, we find that the right-hand side of Eq. \eqref{integralpoissonPP} vanishes, while the left-hand side yields $2 C_q y$, which is not identically zero. However, under typical plasma conditions where $C_q << 1$, and for $y \in [0,1]$, the deviation is negligible, confirming the validity of the approximation.

\section{The case of a statistics of the temperature increments}\label{sec:appendixB}

\begin{figure}
    \centering
    \includegraphics[width=0.99\columnwidth]{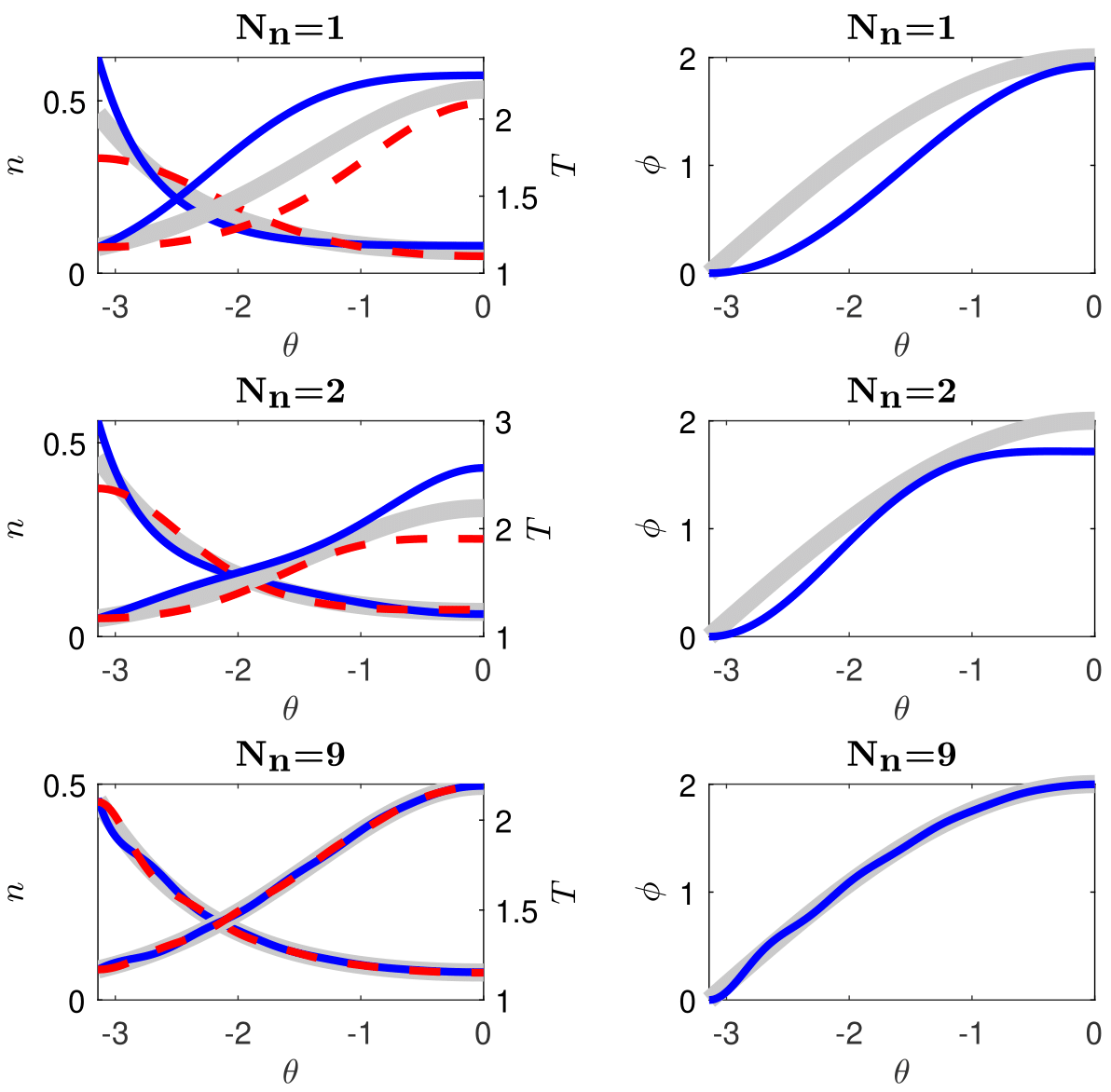}
    \caption{Same quantities of fig. \ref{DensityandPannebitemp} but computed via the statistical distribution of temperature increments given by Eq. \eqref{eq:prob_increments}
    }
    \label{fig:StattempPannemulti}
\end{figure}

In this appendix, we consider the case where the temperature increments follow a statistical distribution. In particular, as previously done in \citep{Barbieri2023temperature, Barbieri2024b, Barbieri2025}, we adopt an exponential distribution for the temperature increments
\begin{equation}\label{eq:prob_increments}
     \gamma(T)=\frac{1}{\langle \Delta T\rangle}e^{-\frac{T-T_0}{\langle \Delta T \rangle}}, \quad T>T_0 .
 \end{equation}
We solve Eq. \eqref{Qnmultitemp} using the distribution $\gamma(T)$ defined above. Once the coefficients $Q_n$ are obtained, we use the same statistical distribution to compute the temperature and density profiles via Eqs. \eqref{NEqtemperaturewithC} and \eqref{NEqdensitywithC}, respectively. Figure \ref{fig:StattempPannemulti} presents the results for the following parameter values: 
$\tilde{g}=2, M=m_p/m_e, C=10^4, \langle \Delta T \rangle=90$. As observed in the isothermal and bitemperature cases discussed in Sections \ref{subsec:thermal} and \ref{subsec:multithermal}, respectively, the electrostatic interaction, when modeled with a sufficiently large number of Fourier modes, produces an electrostatic potential that closely approximates the Pannekoek-Rosseland potential. This approach ensures quasi-neutrality by aligning the temperature and density profiles of electrons and protons across the entire loop.

\begin{acknowledgments}\textbf{Acknowledgments.}
\end{acknowledgments}
The author wishes to thank the anonymous reviewers for their valuable comments, which have helped to improve the presentation of the manuscript.

Valuable discussions with Arnaud Zaslavsky, Simone Landi, Lapo Casetti, Filippo Pantellini and Etienne Berriot are gratefully acknowledged. 

The author acknowledges the Fondazione CR Firenze under the projects \textit{HIPERCRHEL}.

\begin{acknowledgments}\textbf{Funding.}
The author wants to thank the Sorbonne Université in the framework of the Initiative Physique des Infinis for financial support.
\end{acknowledgments}

\begin{acknowledgments}\textbf{Declaration of interests.}
The author reports no conflict of interest.
\end{acknowledgments}

\appendix

%%%%%%%%%%%%%%%%%%%%%%%%%%%%%%%%%%%%%%%%%%%%%%%%%%%%%%%%%%%%%%%%
\bibliographystyle{jpp}
\bibliography{jpp-instructions}
\end{document}